\documentstyle[12pt]{article}
\begin{document}
\title{The $\rho$-Meson as a Collective Excitation\thanks{
       This work was supported
       in part by the grants from CERN No PCERN C/FAE/74/91;
       STRDA/C/CEN/564/92 and Russian Fund of Fundamental Research
       No 94-02-03028.}}

\author{A.A.Osipov\thanks{Permanent address: Joint Institute for
       Nuclear Research, Laboratory of Nuclear Problems, 141980 Dubna,
       Russia.} \\
       Centro de F\'{\i}sica Te\'{o}rica, Departamento de F\'{\i}sica \\
       da Universidade de Coimbra, P-3000 Coimbra, Portugal}

\maketitle
\begin{abstract}
A model of the $\rho$-meson as a collective excitation of
$q\bar{q}$ pairs in a system that obeys the modified
Nambu--Jona-Lasinio Lagrangian is proposed.
The $\rho$ emerges as a dormant Goldstone boson. The origin of the
$\rho$-meson mass is understood as a result of spontaneous chiral
symmetry breaking. The low-energy dynamics of $\rho$, $\pi$,
$\omega$ and $\gamma$ is consistently described in this new framework.
The model accounts for the origin of the celebrated
Kawarabayashi--Suzuki--Riazuddin--Fayyazuddin relation.
\end{abstract}
\newpage

If we assume that pions are bound states of the quark and
antiquark, one can explain the small pion mass by analogy with
superconductivity, considering these ground-state mesons to be
collective excitations of $q\bar{q}$ pairs. Nambu and Jona-Lasinio
(NJL) have done it in their pioneering work \cite{1}. In the NJL model
pions appear as massless Goldstone modes associated with the
dynamical spontaneous breakdown of chiral symmetry of the initial
Lagrangian. An attractive idea is to treat low-lying hadronic
states similarly to the pions as collective excitations of the QCD
vacuum, their masses resulting from dynamical breaking of chiral
symmetry. Is it possible to modify the NJL Lagrangian in such a way
as to simulate the above situation at least for the lowest meson
states, first of all for the $\rho$-meson? An encouraging fact is the
existence of a relationship between the $\rho$-meson and pion
parameters expressed by the well-known
Kawarabayashi--Suzuki--Riazuddin--Fayyazuddin (KSRF) relation \cite{2}.

Another principle underlies the frequently used current models describing
vector particles on the basis of the extended NJL Lagrangian \cite{3}. The
$\rho$-meson mass is not determined by the quark condensate but
fixed {\it ad hoc} by choosing the four-quark interaction constant. This
breaks the integrity of the approach where mass spectra of
collective $q\bar{q}$ pair excitations depend only on the order
parameter.

A new approach to consideration of vector excitations in the
extended NJL model is proposed in the present Letter. It solves the
problem of the $\pi$--$\rho$ system spectrum. The essential
feature of the interpretation of vector mesons which I advocate here is
that they are the result of $(\bar{q}\stackrel{\leftrightarrow}{
\partial_\mu}\lambda_aq)^2$ four-quark interactions.
The special ansatz relates constants of scalar and
vector four-quark vertices. A gap solution appears in the vector
channel. This state is described by an antisymmetrical tensor field
and has a mass $m^2_\rho = 10m^2$, where $m\simeq 242$ MeV is the constituent
quark mass. It can be identified with the $\rho$-meson. The
dynamics of the state is studied. An interesting consequence of the
model is the substantiation of the KSRF relation, which is found here
in the limiting case $m_\pi = 0$. Despite the relationship between
four-quark interaction constants, the model still allows freedom in
choosing the common constant characterizing the coupling
between the vector fields and quarks. It can be used to validate the
hypothesis of universality of vector meson interactions.
The non-Abelian anomaly structure of the effective Lagrangian for pions
and spin-one mesons is different from the
conventional one, obtained by Witten's ``trial-and-error'' gauging
of the Wess-Zumino term in the paper by \"{O}.Kaymakcalan, S.Rajeev
and J.Shechter \cite{4}. The model forbids the $\omega\rho\pi$ vertex
at the quark one-loop level. The $\omega\rightarrow\pi\pi\pi$ decay
with the only contact term is well described.
Certainly it may be related with the new
understanding of the origin of spin-one mesons.

Let us consider the $U(2)$ symmetrical Lagrangian
with four-quark interaction of the Nambu--Jona-Lasinio type
(with the coupling constant $G_1$) and additional
vector interaction (with the coupling constant $H$)
\begin{equation}
L(q)=\bar{q}(i\gamma^\mu\partial_\mu -\widehat{m})q+
     \frac{G_1}{2}\left[(\bar{q}\tau_aq)^2+
                        (\bar{q}i\gamma_5\tau_aq)^2
                  \right]+\frac{H}{2}(\bar{q}\tau_a
     \stackrel{\leftrightarrow}{\partial_\mu}q)^2,
\end{equation}
where $\bar{q}=(\bar{u}, \bar{d})$ are coloured current quark fields
with current mass $\widehat{m}, \tau_a=(\tau_0, \tau_i), \tau_0=I,
\tau_i$ are Pauli matrices of the flavour group
$SU(2)_f$. We represent the four-quark interaction constant with a
derivative as a product of two quantities of the same
dimensionality $H=G_2G$. The dimensionality of the constants is
$[G_\alpha ]=M^{-2}$. The derivative
$\stackrel{\leftrightarrow}{\partial_\mu}$ is defined as
$2\bar{q}\stackrel{\leftrightarrow}{\partial_\mu}q=(\bar{q}\partial_\mu
q-\partial_\mu\bar{q}q)$. We do not require here the chiral symmetry for
vector interactions. This question will be investigated elsewhere.

With the help of auxiliary meson fields $\bar{\sigma}, \tilde{\pi},
\tilde{\rho_\mu}$, Lagrangian (1) takes the form
\begin{equation}
L'=\bar{q}(i\gamma^\mu\partial_\mu -\widehat{m}+\bar{\sigma}+i\gamma_5
   \tilde{\pi}+i\sqrt{G}\tilde{\rho_\mu}\stackrel{\leftrightarrow}{
   \partial_\mu})q-\frac{\bar{\sigma}^2_a+\tilde{\pi}^2_a}{2G_1}+
   \frac{\tilde{\rho}^2_{\mu a}}{2G_2}.
\end{equation}
Here $\bar{\sigma}=\bar{\sigma}_a\tau_a, \tilde{\pi}=\tilde{\pi}_a\tau_a,
\tilde{\rho}_\mu =\tilde{\rho}^a_\mu\tau_a$, and the derivative
$\stackrel{\leftrightarrow}{\partial}_\mu$ affects only
quark fields. Owing to the dimensional constant $\sqrt{G}$, the vector field
will have the same dimensionality as other meson fields. The vacuum
expectation value of the field $\bar{\sigma}_0$ turns out to be different from
zero. To obtain the physical field $\tilde{\sigma}_0$ with
$<\tilde{\sigma}_0>=0$ we perform a field shift resulting in a new quark
mass $m$ to be identified with the mass of the constituent quarks:
$\bar{\sigma}_0-\widehat{m}=\tilde{\sigma}_0-m, \bar{\sigma}_i=
\tilde{\sigma}_i$, where $m$ is deduced from the gap equation.

Let us integrate over the quark fields in the generating functional
associated with Lagrangian (2). Evaluating the resulting quark
determinant by a loop expansion we get
\begin{equation}
L_{coll}=-i{\rm Tr}\ln\left[1+
         \frac{(\tilde{\sigma}+i\gamma_5\tilde{\pi}+
                i\sqrt{G}\tilde{\rho}_\mu
         \stackrel{\leftrightarrow}{\partial_\mu})}{
         (i\gamma^\mu\partial_\mu -m)}\right]_\Lambda
         -\frac{\bar{\sigma}^2_a+\tilde{\pi}^2_a}{2G_1}+
          \frac{\tilde{\rho}^2_{\mu a}}{2G_2}.
\end{equation}
The index $\Lambda$ indicates that a regularization of the
divergent loop integrals is introduced. In this case it is the
Pauli--Villars regularization \cite{5}. It conserves vector gauge
invariance and at the same time reproduces the quark condensate and
the current quark mass. The Pauli--Villars cut-off $\Lambda$ is
introduced by replacements $\exp (-izm^2)\rightarrow R(z)$ and
$m^2\exp (-izm^2)\rightarrow iR'(z)$, where $R(z)$ is the
regularizing function that satisfies the conditions $R(0) = R'(0) =
\ldots = 0.$

The effective Lagrangian (3) allows us to calculate the leading
low-energy behaviour of any N-point meson function. This
behaviour is governed by the first terms in the expansion of
derivatives \cite{6}. From the requirement for the terms linear in
$\tilde{\sigma}$ to vanish we get a gap equation
\begin{equation}
m-\widehat{m}=8mG_1I_1
\end{equation}
(see Eq.(6) for notation). The effective meson Lagrangian describing
scalars and pseudoscalars is well known \cite{7}, so we turn to the effective
Lagrangian of the vector fields. To the order of $p^4$ the two-point
function $<0|T\tilde{\rho}_\mu\tilde{\rho}_\nu |0>=T_{\mu\nu}$ is given by
\begin{equation}
T_{\mu\nu}(p)=g_{\mu\nu}(G_2^{-1}-GI_0)+\frac{GI_2}{3}(p_\mu p_\nu -
  p^2g_{\mu\nu})(p^2-10m^2)+{\cal O}(p^6).
\end{equation}
Here we use the notation
\begin{equation}
I_\alpha=-\frac{3c_\alpha}{16\pi^2}\int_{0}^{\infty}\frac{dz}{
         z^{3-\alpha}}R(z),
\end{equation}
where $\alpha =0,\,1,\,2, c_0=20, c_1=i, c_2=-1$.

Let us introduce the ansatz $1-HI_0 = 0$. On the one hand, it ensures
gradient invariance of self-energy (5) and on the other it relates
$G_1$ and $H$ constants by the condition $HI_0=8G_1I_1+\widehat{m}/m$
via gap equation (4). It was stressed by T.Eguchi \cite{8} that one
can create a local gauge symmetry starting from global invariance
in nonlinear spinor theories. In Lagrangian (2) we have a
collective excitation $\tilde{\rho}^a_\mu$ which is coupled to a
conserved current $(\bar{q}\stackrel{\leftrightarrow}{\partial}_\mu
\tau_aq)$. Hence (2) is invariant under $\tilde{\rho}\rightarrow
\tilde{\rho}+\partial\alpha$ with an arbitrary $\alpha (x)$. Although the
term $\tilde{\rho}^2_{\mu a}/2G_2$ seems to spoil this invariance, it
in fact eliminates the gauge-noninvariant part coming from radiative
corrections and preserves the gauge invariance.

Renormalizing the $\rho$-meson field $\tilde{\rho}=\sqrt{3/I_2}\rho$
we arrive at the Lagrangian
\begin{equation}
L_\rho =G\left(-\frac{1}{2}\partial_\mu\rho^a_{\mu\nu}
                           \partial_\lambda\rho^a_{\lambda\nu}
               +\frac{1}{4}m^2_\rho\rho^a_{\mu\nu}\rho^a_{\mu\nu}
         \right),
\end{equation}
where the $\rho$-meson mass is expressed in terms of the
constituent quark mass
\begin{equation}
m^2_\rho =10m^2.
\end{equation}
Thus, like the mass of the scalar field $\sigma$, the $\rho$-meson mass
results from spontaneous breaking of chiral symmetry. This
collective state is described by the antisymmetrical tensor field
$\rho_{\mu\nu}$. It is interesting that J.Gasser and H.Leutwyler used the very
Lagrangian to describe the $\rho$-meson while estimating the effect
of resonance states on the low-energy structure of Green functions
\cite{9} (see also \cite{10}). The wave equation associated with
Lagrangian (7) has the form
\begin{equation}
\left\{
\begin{array}{ll}
&\dot{\Pi}_i+\partial_i\partial_j\rho_{j0}-m^2_\rho\rho_{0i}=0 \\
&\partial_j\Pi_i-\partial_i\Pi_j-m^2_\rho\rho_{ji}=0,
\end{array}
\right.
\end{equation}
where the canonical momentum $\Pi_i$ is $\Pi_i=-\partial_\mu\rho_{\mu i}
(i,\,j = 1,\,2,\,3).$ According to these equations, only the $\rho_{0i}$
field components oscillate. The components $\rho_{ij}$ are frozen. At
$m_\rho = 0$ (or $m=0$) equation (9) possesses symmetry under the
following transformations: $\rho_{i0}\rightarrow\rho_{i0}+\partial_i
\theta (\vec{x}), \vec{\nabla}^2\theta (\vec{x})=0.$ One can use this
freedom imposing the gauge fixing requirement $\partial_i\rho_{i0}=0$.
In this case the field keeps only transverse degrees of freedom,
and the vector mode is fully frozen. It means that if $F(q^2)$ is the
vector meson coupling constant for virtual momentum $q_\mu$ then it
reduces to zero in the limit $q^2 \rightarrow 0$ (or $m \rightarrow 0$).
Decoupling of the vector field occurs. D.Caldi and H.Pagels already
supposed the existence of this dormant state earlier \cite{11}.

To proceed, we must fix the parameters of the model. For direct
comparison with the empirical numbers we have used the values for the
pion decay constant, $f_\pi$, and the pion mass, $m_\pi$, close to their
physical values, $f_\pi\simeq 93$ MeV and $m_\pi\simeq 139$ MeV,
respectively. Within the purely fermionic NJL model \cite{12}, a similar
set of parameters has already been determined. For the case at hand,
we used $G_1=7.74\,\mbox{GeV}^{-2}, \Lambda =1.0$ GeV and $\widehat{m}=
5.5$ MeV. With these parameters, we find $f_\pi =93.1\,
\mbox{MeV}, m_\pi =139\,\mbox{MeV}, m=241.8\,\mbox{MeV}$ and $m_\rho =
765$ MeV. Let us note that we should not introduce the new parameters
to describe the mass of the vector particles.

Let us discuss the main features of the $(\rho , \omega )$--$(\pi , \gamma )$
interactions in the model under consideration. As far as calculations
are concerned, we shall not go into detail leaving it to a future
lengthier paper. The relevant effective Lagrangian contains the
following terms
\begin{eqnarray}
L_I&=&\frac{e}{f_\rho}\partial_\mu A_\nu (\rho_{\mu\nu}+\frac{1}{3}
      \omega_{\mu\nu})+f_{\rho\pi\pi}\varepsilon_{ijk}\rho^i_\mu
      \pi^j\partial_\mu\pi^k-\frac{ef_{\rho\pi\gamma}}{2f_\rho f_\pi}
      \varepsilon^{\mu\nu\lambda\sigma}\partial_\mu A_\nu
      \partial_\lambda\rho_\sigma^i\pi^i \nonumber \\
   & &-\frac{f_{\omega 3\pi}}{
      f_\rho f_\pi^3}\varepsilon^{\mu\nu\lambda\sigma}
      \varepsilon_{ijk}\omega_\mu\partial_\nu\pi^i
      \partial_\lambda\pi^j\partial_\sigma\pi^k+\ldots ,
\end{eqnarray}
where we have introduced a symbol $A_\mu$ for an electromagnetic
field\footnote{It is traditional to include the interaction with the
electromagnetic field via the standard replacement of the derivative
$\partial_\mu$ in Lagrangian (3) by the covariant $\partial_\mu - ieQA_\mu$.}
and left out pieces of no concern to us here.

The constants of the $\rho\rightarrow\pi\pi$ and
$\rho\rightarrow\gamma$ transitions have the form
\begin{equation}
f_{\rho\pi\pi}=\frac{\sqrt{3G}m^2(m^2_\rho -2m^2_\pi )}{4\pi^2f_\pi^3},
\,\,\,\,\,\,\,\frac{1}{f_\rho}=\frac{\sqrt{G}f_\pi}{\sqrt{3}}.
\end{equation}
Hence we derive the relation
\begin{equation}
m^2_\rho -2m^2_\pi =\frac{4\pi^2f_\pi^4}{3m^2}f_{\rho\pi\pi}f_\rho
\end{equation}
which generalizes the remarkable KSRF result $m^2_\rho =2f_{\rho\pi\pi}
f_\rho f^2_\pi$ to the case of broken chiral symmetry $\widehat{m}\neq 0$.
Ignoring the pion mass, one understands the origin of the mysterious two
\begin{equation}
2\sim\frac{(2\pi f_\pi)^2}{3m^2}=1.96.
\end{equation}
Relation (11) can be used to deduce
\begin{equation}
\frac{f_{\rho\pi\pi}f_\rho}{4\pi}=\frac{3m^2(5m^2-m^2_\pi)}{8\pi^3
      f^4_\pi}=2.55
\end{equation}
The available data \cite{13} require $f^2_{\rho\pi\pi}/4\pi = 2.9$
from $\Gamma(\rho\rightarrow\pi\pi)=149$ MeV and $f^2_\rho /4\pi =2.0$
from $\Gamma(\rho\rightarrow e^+e^-) = 6.8$ keV. Combining these data
we obtain $(f_{\rho\pi\pi}f_\rho /4\pi )^{exp}=2.41.$ Agreement in (14)
is even better because the $\rho e^+e^-$ coupling $f^2_\rho /4\pi$ is
corrected from 2.0 to 2.4.

The dimensional parameter $\sqrt{G}$ can play a special role. Taking
\begin{equation}
\sqrt{G}=\frac{\sqrt{2}\pi f_\pi}{m\sqrt{5m^2-m^2_\pi}}\simeq
         \sqrt{\frac{2}{5}}\frac{\pi f_\pi}{m^2}=0.62\,\mbox{fm},
\end{equation}
we get the universality condition $f_{\rho\pi\pi}=f_\rho$. The
slight discrepancy observed in $f_{\rho\pi\pi}$ and $f_\rho$ can
be attributed to deviation from the value (15). It should be pointed
out that the value of this parameter has no influence on the form of
the KSRF relation. This is different from the hidden local symmetry
approach \cite{14}, where the requirement of universality leads to
the KSRF relation and {\it vice versa}.

We also note that the coupling constant $f_{\rho\pi\gamma}$ is equal to
\begin{equation}
f_{\rho\pi\gamma}=\frac{m^2_\rho +m^2_\pi}{8\pi^2 f^2_\pi}.
\end{equation}
Therefore the leading coupling that allows these transitions is
${\cal O}(p^4)$. This fact is known from the paper \cite{10}.

We have a very interesting situation with the $\omega\rightarrow\pi\pi\pi$
decay. An early investigation, published in 1962 by M.Gell-Mann,
D.Sharp and W.Wagner (GSW) \cite{15}, took the point of view that
the $\rho$-pole term completely dominates in this process. In our model
there is no coupling that would induce this transition at the quark one-loop
level. In this leading approximation the $\omega\rho\pi$ vertex is
equal to zero and can appear only at the level of mesonic loops at
higher order of momentum expansion. Instead of the GSW model, we can
describe this decay allowing for the only contact term in
$\omega\rightarrow 3\pi$. The constant value of $f_{\omega 3\pi}$
gives good agreement with the Dalitz plot of these decays.
Our calculations of the box diagrams show that
\begin{equation}
f_{\omega 3\pi}=\frac{3m^2}{4\pi^2 f^2_\pi}\left(1+
      \frac{m^2_\omega +3m^2_\pi}{12m^2}\right).
\end{equation}
Therefore, the resulting width $\Gamma (\omega\rightarrow\pi^+\pi^0
\pi^-)=6.0$ MeV is in satisfactory agreement with the experimental
value $7.49\pm 0.09\pm 0.05$ MeV \cite{13}.

A final remark:
The proposed model indicates that vector mesons, like pseudoscalar
ones, are collective excitations of the quark sea. This statement
is grounded on two significant consequences of the approach in
question. One is a surprisingly good description of the
$\rho$-meson mass, the other is the KSRF relation which arises naturally
here and whose character has remained obscure so far.

\end{document}